# Developing a Security Testbed for Industrial Internet of Things

*Muna AL-Hawawreh, Elena Sitnikova*

**Abstract—** While achieving security for Industrial Internet of Things (IIoT) is a critical and non-trivial task, more attention is required for brownfield IIoT systems. This is a consequence of long life cycles of their legacy devices which were initially designed without considering security and IoT connectivity, but they are now becoming more connected and integrated with emerging IoT technologies and messaging communication protocols. Deploying today's methodologies and solutions in brownfield IIoT systems is not viable, as security solutions must co-exist and fit these systems' requirements. This necessitates a realistic standardized IIoT testbed that can be used as an optimal format to measure the credibility of security solutions of IIoT networks, analyze IIoT attack landscapes and extract threat intelligence. Developing a testbed for brownfield IIoT systems is considered a significant challenge as these systems are comprised of legacy, heterogeneous devices, communication layers and applications that need to be implemented holistically to achieve high fidelity. In this paper, we propose a new generic end-to-end IIoT security testbed, with a particular focus on the brownfield system and provide details of the testbed's architectural design and the implementation process. The proposed testbed can be easily reproduced and reconfigured to support the testing activities of new processes and various security scenarios. The proposed testbed operation is demonstrated on different connected devices, communication protocols and applications. The experiments demonstrate that this testbed is effective in terms of its operation and security testing. A comparison with existing testbeds, including a table of features is provided.

## I. INTRODUCTION

THE Industrial Internet of Things (IIoT) is a special case of IoT paradigms in which machines, computers and people can enable intelligent industrial operations based on advanced data analytics. Industrial systems are becoming increasingly capable of working automatically and intelligently and even responding to unpredicted events (e.g., machine's failure). However, there is a security risk associated with connecting these critical industrial systems to the IoT technologies [1]. A greater concern relates to 'brownfield' systems in which existing legacy industrial systems such as Programmable Logic Controller (PLC), Remote Terminal Unit (RTU), Supervisory Control and Data Acquisition (SCADA), Input/Output (I/O) devices and among others interoperate with IoT technologies [2]. These systems, such as those of energy, water, buildings, roads, and factories have been around for decades and were designed to have long service lives without considering their connectivity and security [3]. Replacing them with new gadgets and devices designed to be secure from scratch (i.e., 'greenfield' systems) is not technically or economically feasible [4, 5]. Furthermore, most of today's security solutions and methodologies are not applicable for such implementation because they are Information Technology (IT)-centric and do not take into account a system's safety, resilience and reliability [6-8]. As a consequence, there is a security-gap in brownfield systems which raise the need to infuse these systems with new security models to protect them and mitigate potential risks.

To develop new, efficient and holistic solutions, more research should be conducted. However, this is often restricted by a lack of realistic data about a system's communications and activities, as well as potential cyber-attacks [9-11]. Unfortunately, it is extremely difficult to obtain realistic data from an actual environment for security and privacy reasons which may lead researchers to make inexact assumptions and limit the applicability of their results. Therefore, realistic data and system test models are essential for researchers to be able to understand the current threats landscape and attack techniques in brownfield IIoT systems with rapidly emerging new devices and connectivity protocols. Existing IoT testbeds cannot be used to test IIoT systems' security (in particular brownfield) as these industrial systems have special requirements such as safety, resilience and reliability, and the need for the integration between legacy and new technologies [4, 12-14]. Thus, it is critical to offer researchers with accurate and realistic testbeds to optimize ongoing research and help the research community to validate its security hypotheses. To achieve high levels of fidelity, there is a need for a holistic, end-to-end IIoT testbed that is generic, accurate, relatively inexpensive and easy to reproduce.

This paper proposes a new IIoT testbed for testing and analyzing security issues related to a system's components (e.g., data, communication, and devices). The main contributions of this paper are thus as follows:

1) We propose a new generic end-to-end IIoT security testbed, with particular emphasis on brownfield implementations. We call it Brown-IIoTbed.
2) We develop a Brown-IIoTbed with free open sources

M. AL-Hawawreh, and E. Sitnikova are with school of engineering and information technology, University of New South Wales (UNSW) - Australian Defence Force Academy (ADFA), Canberra, Australia.(e-mail: m.al-hawawreh@student.adfa.edu.au).

software and cost-affordable hardware.
3) We investigate and test various security threats based on the STRIDE model, as well as security evasion/reverse shell backdoor against router/firewall to demonstrate the feasibility of Brown-IIoTbed.
4) We also provide machine learning approaches for intrusion detection and introduce an example of malicious payload hunting and intelligence as a proactive defense approach (early detection).
5) We provide a comprehensive analysis and comparison of existing testbeds with our Brown-IIoTbed.
6) We release all the implementation of testbed into the GitHub[1], so the researcher can easily reproduce it.

The remainder of this paper is organized as follows: related work in section II; a description of a generic IIoT system architecture in section III; development and implementation of the testbed (Brown-IIoTbed) in section IV; security testing and analysis using Brown-IIoTbed in section V; comparison with existing testbeds and a discussion in section VI; and the conclusion and future work in section VII.

## II. RELATED WORK

Several IoT/IIoT testbeds have been presented by researchers to validate their hypotheses related to various issues in these systems. Patel *et al.*[15] investigated the challenge of designing and experimenting robust IoT applications such as smart traffic and lighting. In their testbed, the authors utilized a Message Queuing Telemetry Transport (MQTT) protocol and Kafka-software platform to transfer data and visualize it in a dashboard to provide knowledge about these applications. In related work, Choosri *et al.* [16] presented a traffic management system testbed to investigate how IoT technologies can be applied to solve the practical requirements for human-oriented traffic control. They used a Java-based application to control Radio Frequency Identification (RFID) readers tagged to vehicles. Another testbed was developed by Deshpande, Pitale, and Sanap [17] for detecting abnormal signals in an industrial automation system. They used a set of sensors (e.g., temperature, and pressure) that sent analog signals to an Android smartphone using Bluetooth. Similarly, Merchant and Ahire [18] introduced a simple industrial automation system testbed for monitoring the blade-aging system of a cutter tool. These presented testbeds did not provide a complete IIoT system and were designed for specific use.

Furthermore, few labs globally have developed large-scale open IoT/IIoT testbeds to test new devices, applications and technologies. For example, Industrial Internet Consortium (IIC) [19] introduced 'INFINITE' as an innovation for building a software-defined infrastructure testbed to drive the growth of IIoT and facilitate a process for testing new applications. 'INFINITE' involved multiple platforms across mobile, cloud, sensor and analytics. It was deployed in Ireland and included multiple IIoT service providers. The Federated Interoperable Semantic Testbed and Applications (FIESTA)-IoT [20] provided large-scale experimental infrastructure for heterogeneous IoT technologies through 10 testbeds distributed around the world. It included smart Santander for a smart city, smart Institute of Communication Systems (ICS) for a smart building, Sound City (SC) for collecting data from smart mobiles, KETI, ADREAM, FINE and Network Implementation Testbed Open Source (NITOS) for smart buildings, EXTEND for sea and underwater environments, Tera.4Agri for smart agriculture and Real Data Center (RDC) for the energy consumption of data center solutions. These testbeds focused on collecting various data types based on different communication technologies, including Wi-Fi, LTE, WIMAX, Zigbee, and 4G/5G terminals. Another open IoT testbed is Japan-wide Orchestrated Smart/Sensor Environment (JOSE) [21]. JOSE focused on collecting data from multiple wireless networks, each of which represented a specific IoT service and used virtual machines for its storage and computations. Also, various communication protocols were used to connect sensors and gateways with middleware and virtual networks including wireless LAN, LTE, and 3G. All these discussed testbeds concentrated only on testing applications and collecting IoT data, and most did not provide clear descriptions of their configurations and components.

Although security is one of the major challenges that IoT/IIoT deployments encounter, few studies have focused on IoT/IIoT security testbed. For example, Siboni *et al.* [22] presented a security testbed capable of testing various IoT devices and physical access media, including Wi-Fi, Zigbee, and Bluetooth. They tested various security issues such as discovering IoT devices' vulnerabilities, detecting anomalies using machine learning, and evaluating testbed's resilience against denial of service (DoS) attacks. However, their proposed testbed focused only on IoT devices and specific physical access media. It also provided an industrial IoT scenario, in particular, a closed control loop, using simulators only. Berhanu, Abie, and Hamdi [23] presented security testbed for smart health applications. In their experiment, data was collected from multiple sensors via smartphones, sent to a storage device and then displayed for end-users through multiple interfaces. This testbed focused on validating energy consumption rather than security issues and presented only a specific use. Moreover, Hossain *et al.* [24] introduced a distributed testbed providers and multiple users which could be rented by end-users to deploy their devices or conduct IoT research experiments (e.g., security testing). The authors argued that, as their testbed was designed based on the principle of cloud services, it provided these services through users registering for allocation of resources which were managed and monitored using the management components. The testbed was evaluated based on a process of allocating various IoT-simulated devices, with the results showing that it could provide a reasonable and manageable performance despite some clients possibly facing a short delay in reserving their demand for resources. However, as evaluating its performance based only on simulated devices could not represent those of real systems and their conditions, this prototype could not be considered a real service.

---
[1] https://github.com/Alhawawreh/Brown-IIoTbed.

In summary, although there is a great deal of literature available on developing testbeds for IoT and IIoT systems with different objectives, most existing testbeds focus on collecting data using various sensor types deployed in wireless sensor networks and supporting different networking wireless communications while others depend on simulators to model the data of IoT/IIoT devices and some involved specific target applications. Open testbeds and those presented as service prototypes are confronted by the challenge of authenticating valid users and offering only restricted access to some collected data due to privacy concerns. They are also highly complex and cannot be adapted to satisfy users' requirements for security testing. Almost all of the existing testbeds cannot guarantee the fidelity of IIoT systems as they do not follow a standard architecture model. Also, the IoT implementations cannot be used to test security issues related to industrial systems since IIoT systems have special requirements regarding safety, resilience, and reliability. The interoperability using various messaging communication protocols, and the integration between legacy control systems and IoT technologies have not been investigated. A holistic end-to-end testbed has not been presented for IIoT systems, in particular brownfield systems, for security testing. Therefore, as there are critical gaps, it is necessary to develop a testbed that focuses on providing a simple and accurate end-to-end IIoT testbed with a high level of fidelity. This paper introduces a new end-to-end IIoT security testbed with a particular focus on recent IIoT connectivity protocols, interoperability-supportive devices, and brownfield systems because they are the most essential for obtaining new security solutions as well as a testbed for evaluating those solutions.

### III. DESCRIPTION OF GENERIC IIoT SYSTEM ARCHITECTURE

An Industrial Control System (ICS) usually uses the Purdue Enterprise Reference Architecture (PERA) as the reference architectural model. Its main concept is based on dividing enterprise and ICSs into vertical and hierarchal segments, including fieldbus, control bus, and corporate, that function in a similar way [25]. Arguably, this model is still adopted in terms of its functionality for providing downward and upward information flows among a brownfield IIoT system's levels. However, because of the implementation of new digital networks and various technologies, such as mobile, edge, and cloud computing, as well as changes in information and command flows through these systems, new architectural reference models have been developed for IIoT systems.

The Industrial Internet Consortium (IIC) published an architectural model for IIoT system applications, called the Industrial Internet Reference Architecture (IIRA), that helps in understanding the implementation of greenfield and brownfield IIoT systems in real environments [26]. It presents different descriptions for these implementations as it is generally case-specific. However, from a variety of such architectures, we extract a generically designed model that can represent the main implementation and functionality of IIoT systems, a horizontal modular with the following three basic tiers.

**Tier1:** The edge tier consists of physical assets/filed devices that form a closed control loop with a senor, actuator, and controller and an edge gateway that provides real-time data analytics, storage, and control. It also connects a physical system with other digital ones and may include an optional fog node for performing various real-time operations.

**Tier2:** The platform tier receives, processes, and relays control commands from the enterprise tier to the edge one. It provides data analytics, storage, and management services that could be implemented in corporate or cloud data centers.

**Tier3:** The enterprise tier represents a service network for an IIoT system and includes an Application Programming Interface (API), Web-SCADA and Human User Interface (HUI) to enable human interactions with applications, the issuance of commands to the edge tier, the making of smarter decisions and the performance of maintenance.

The IIRA model focuses totally on the basic characteristics and cross-cutting features of any IIoT system implementation and can be described as follows. Firstly, it emphasizes a closed control loop, which is clearly defined at the edge tier, for collecting and analyzing data and controlling the system. Secondly, it has a large-scale closed control loop in an IIoT system that includes all the system's tiers (i.e., edge, platform, and enterprise) and uses the data collected from the physical control systems for analysis at the corporate/cloud data center servers. Then, it makes a smart decision regarding the operational process and eventually concentrates on influencing the systems in the edge tier. Finally, it also emphasizes the interoperability features in an IIoT implementation; for example, the interoperability between various messaging communication protocols, networks, or types of devices and the interoperability among systems so that a one-IIoT system could use the cloud data collected from different IIoT systems.

Given that, we use the IIRA as a reference model to build a more realistic testbed that helps to demonstrate the fidelity and credibility of security research. However, the horizontal interoperability is out of the scope of this paper as our focus on providing simplified IIoT testbed for security testing rather than the system of systems integration.

### IV. DESIGN AND DEVELOPMENT OF THE TESTBED

*A. Description of generalized Brown-IIoTbed architecture*

Fig.1 illustrates the architecture of Brown-IIoTbed based on IIRA model, it is a generalized prototype which means that any type of sensor, actuator, and industrial control device (such as PLC, RTU, and I/O devices), can be involved in it. Also, to provide more flexible capabilities for performing various IIoT applications and security tests, any relevant simulator can be used as a physical system as well as various human interfaces, including laptops, Personal Computers (PC), smartphones and tablets, to provide interactions with edge devices and visualizations of data analytics. The testbed is divided into the following zones.

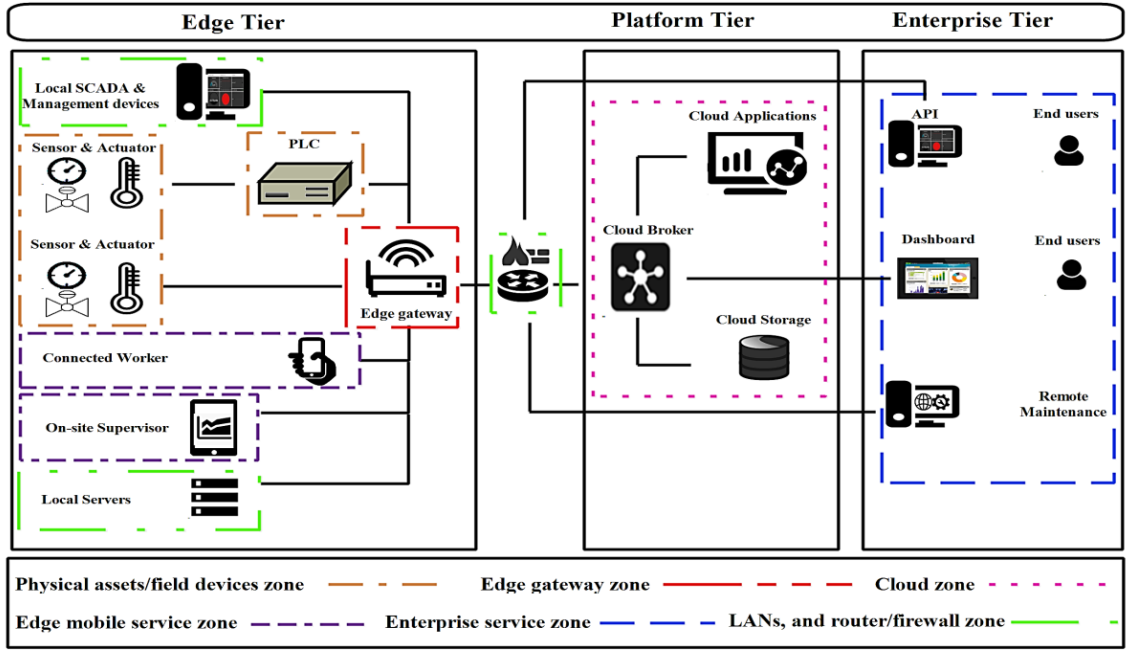

Fig. 1. Testbed architecture based on IIRA model

**Physical assets/filed devices zone**: Whenever a physical assets/filed devices are represented, the device and/or simulator deployed should behave similarly to the actual system and have a deterministic scan cycle, response time and accurate response for each of the inputs provided. The cyber-physical links and access medium between these devices and other system components should represent the actual medium and communications of the systems deployed. Therefore, various industrial physical assets/filed devices are used, each of which achieves the closed control loop required by the IIRA model. Firstly, a PLC, a master device that uses the MODBUS protocol, is connected to an I/O slave device represented by a physical analog sensor and actuator. The PLC device is also connected using the MODBUS/TCP protocol to the edge gateway. Secondly, two I/O slave devices with two different serial fieldbus protocols are used. They represent digital sensors and actuators that are controlled directly by the edge gateway as a master device. Finally, a simulator (our developed JavaScript script) acts as the sensor, controller and actuator, is used.

**Edge gateway zone:** An edge gateway plays a critical role in supporting Machine-to-Machine (M2M) communication, collecting data from various physical devices, storing it as a time-series database (local data historian) and sending it to the cloud. It also acts as a microcontroller for I/O devices and a master device for PLC at the same time as well as supporting the interoperability and integration of various communication protocols and different legacy systems. It can also provide a local SCADA or API system as a web service for monitoring and controlling these physical processes and devices [27, 28].

Edge mobile service zone: A mobile device runs at the edge tier to provide Human-to- Machine (H2M), and Machine-to-Human (M2H) communications. It represents the smartphones and tablets used by connected workers and their on-site supervisor. The connected workers use a mobile application to wirelessly access the physical devices using the edge gateway to read data, send commands and check the status of the actuators as well as perform specific Python scripts for maintenance after receiving a notification from edge gateway. The on-site supervisor can use the tablet as a web interface device to access the statistical figures and data analytics (i.e., as a mobile SCADA clients).

**Local Area Networks (LANs), router and firewall zone:** The edge gateway connects with local networks that provide local SCADA and various simple IT services in both wired and wireless forms. For example, it connects with a mail server to send e-mail notifications to connected workers and with an IT PC to provide local management and maintenance for edge devices. A router is used to connect between the edge systems and external world i.e., Wide Area Network (WAN), this router is also with firewall capabilities to achieve a high-fidelity security testing as the edge physical devices and systems cannot connect directly with the Internet without accessing a control list.

**Cloud zone:** The cloud platform consists of a broker that connects with the edge gateway to receive the measurements and states of the physical assets/filed devices and a data historian to store the values that fall outside the predefined deadband, with only the significant data sent to the cloud to reduce the amount of bandwidth used and provide real-time processing at the edge gateway [28]. However, Brown-IIoTbed presents various scenarios for collecting data, including the deadband and deterministic values, stored in a usable and searchable way, i.e., in an indexed database, to allow analyses or machine learning processes to be performed in order to make further decisions. The cloud application provides the analytics results and statistical data graphs as a service. In fact, it helps to transform the raw data into the valuable information sought to satisfy the requirements of other IIoT applications.

**Enterprise service zone**: one of the main objectives of an IIoT system is to provide remote monitoring and control. Therefore, the development and deployment of APIs or Web-SCADA interfaces to provide remote monitoring and real-time data visualization, such as of sensor data, actuator status, current set-point values, statistics and graphs. These interfaces can be accessed using any web browser via any computer or mobile device. Another task that could be performed at this zone is the remote technical maintenance of edge physical devices by control systems' individuals.

*B. System components and structure of Brown-IIoTbed*

The testbed environment illustrated in Fig. 2 shows the hardware, software and communication protocols used in Brown-IIoTbed implementation:

**1) System hardware components**: At the edge tier, two Raspberry Pi and one Arduino mega 2560 devices are used to act as an edge gateway, PLC, and I/O device respectively. These electronic devices are very common IoT devices with multiple I/O pins, are easy to use and with affordable prices. Also, one analog and one digital sensor, i.e., TMP36 (−40°C to +125°C) and DS18B20 (-55°C to +125°C) provide various temperature readings value. The focus is on performing a closed-loop temperature controller which is considered a part of most existing ICSs [52, 53]. To provide a variety of measurement readings, a pressure sensor (MPL3115A2) is used to obtain both pressure (20 to 110 kPa) and temperature values (–40 °C to 85 °C). Various Light Emitting Diode (LED) devices are deployed as the actuators represent the controlling parameters that can influence the closed control loop, such as its switch on/off pump relay or valve. Also, mobile devices, such as iPhones and Android tablets, are integrated into this testbed to provide connectivity with the edge gateway. LANs are represented by laptop and PC devices connected to the edge gateway by a Wi-Fi access point and physical switch (i.e., a 2810-24G managed HP Ethernet switch) respectively. To establish a WAN and external world, an HP server provides various devices and a router/firewall.

**2) System software components:** Various open-source software tools freely available online are used to perform specific tasks. An OpenPLC server runs on a Raspberry Pi B+ device and connects to an OpenPLC software (C++ code) slave that runs on an Arduino Mega 2560 device. The OpenPLC [29] is an open-source fully functional PLC that supports MODBUS/TCP connectivity and ladder logic as a programming language for process control. A simple ladder logic program is created to read the values from an analog sensor, process them based on a function block, and react by sending a command to the actuator. A Node-Red [30] application is used to perform the functionalities of the edge gateway. It is a programming tool for wiring together hardware devices, APIs, and online services with multiple libraries for connecting physical devices, such as a MODBUS PLC, digital sensor, LED, and many more. Multiple Node-Red flows (JavaScript codes) are created to perform the edge gateway's functionality, connect the PLC, I/O devices and other physical assets to the edge gateway, and then the external world. Also, additional flows are created to store real-time data (data historian) and analyze it, control I/O devices, and simulate a closed control loop to provide different limited-range random values of temperature, pressure and humidity measurements. The Node-Red application also helps to establish an API and Web-SCADA. The edge gateway runs txThing Constrained Application Protocol (CoAP) [31] (building based on twisted Python library, as one of the most common CoAP Python framework), Apache2 web server, MYSQL database, Secure Shell (SSH), Dnsmasq (for providing Domain Name Service (DNS), Dynamic Host Configuration Protocol (DHCP), router advertisement and network boot) and Host access point daemon (Hostapd) to provide access points for wireless mobile devices and other services for end-users as well as physical control devices. VMware workstation pro [32], a virtualization tool, is used to create multiple virtual machines at a Windows laptop and PC. The laptop provides a mail service and CoAP client scripts. The hMailserver [33] is run on a Windows virtual machine to provide SMTP mail services, with Python scripts used on another Windows virtual machine to simulate CoAP client behavior. The virtual machines on the PC provide management (Windows) for various services and internal attack tools (i.e., kali Linux).

The iPhone mobile devices use three applications installed from the App Store, the "CoAPClient" app for connecting to the CoAP server, Shortcuts app for creating repeated scripts over SSH and Mail app for receiving notifications from the edge gateway as e-mail. The tablet device runs on a Chrome web browser to access Web-SCADA and the ESXi 6.7 hypervisor [32] is used for virtualization. Multiple virtual machines are created to act as a virtual router/firewall with pfsense software [34], a cloud server running a mosquito MQTT broker [35] to receive data from the edge, MYSQL storing the received data in an indexed database, and a shiny server with the R language [36] creating a cloud data analytics application while other virtual machines act as an API, Web-SCADA, remote maintenance and external attacker tool (i.e., Kali Linux). The virtualization technique is adopted since it is often inexpensive, reduces the amount of equipment required and simplifies the physical compositions of systems. The iMacro [37] is used to automate the behavior of human on web interfaces, and Python scripts for automating SQL clients.

**3) Communications and physical links:** In this implementation, the heterogeneous application layer (messaging) protocols concentrating mainly on those that are common in IIoT are used. For physical media, both wired and wireless communication (Ethernet IEEE 802.3 and Wi-Fi IEEE 802.11 respectively) are employed, with the PLC, PC and WAN devices configured to use the former and the mail server, CoAP simulator and mobile devices the latter. The edge gateway is configured to use both. The legacy industrial protocol MODBUS/TCP, the most common industrial protocol in brownfield implementations, is used to connect the PLC device to the edge gateway and the MODBUS serial to connect the Arduino/ slave device to the PLC, i.e., the Raspberry Pi B+/master. The fieldbus serial communications I2C and 1-wire connect the sensor and actuators to the edge gateway as I/O devices. The DNS and CoAP protocols enable a mobile device to connect to the CoAP server and access point at the edge gateway while the SSH protocol is used in most devices deployed to perform remote access and management. The MQTT protocol, the second common protocol after HTTP in IoT, publishes data to the cloud server and WebSocket accesses APIs from mobile devices and the local API-edge gateway monitor. The HTTP protocol is used for Web-SCADA from an external network (WAN), SMTP, and IMPA ones to send and receive e-mail notifications and SQL to access a local data historian.

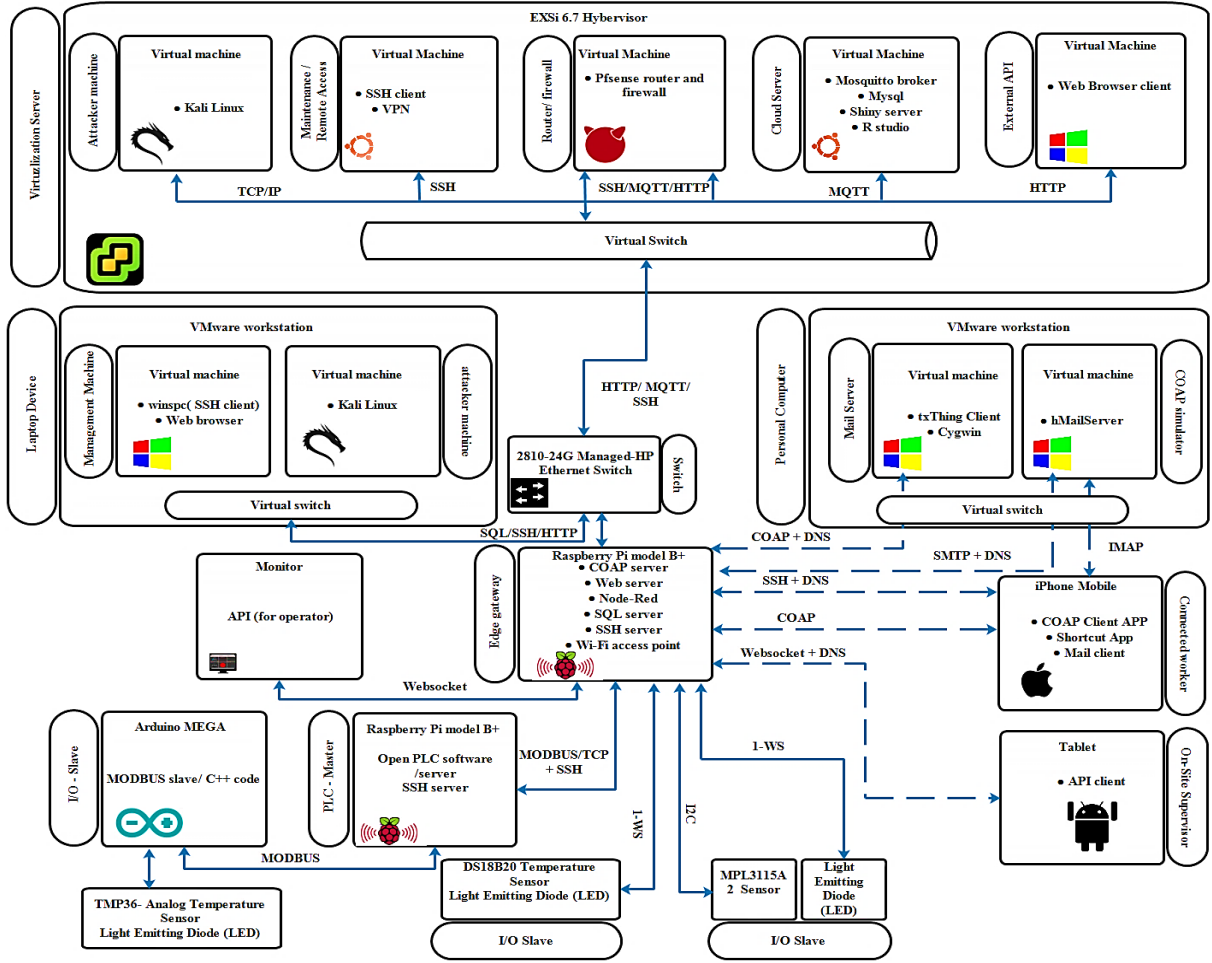

Fig. 2. Structure of the Brown-IIoTbed

*C. Cases using Brown-IIoTbed functionalities*

Uses of the proposed testbed are discussed in the following examples of cases of generalized and non-specific systems. They concern the complete proposed system and are aimed at illustrating the feasibility of our proposed testbed, Brown-IIoTbed, as our focus is on hardware and software implementations for security testing:

**Case1. Quick checkup and response using mobile application:** A connected worker uses his/her mobile application to send a request to the edge gateway over the CoAP protocol-wireless network, asking for the current reading value of the I2C sensor. He/she can also send a command to switch on/off the actuator, that is, the I-wire LED at any time.

**Case 2. Notification and warning via e-mail:** The edge gateway connects to the local mail server to send an email notification via the SMTP protocol to a connected worker's mobile. This email confirms that a specific command has been sent to the actuator or initiates a warning that the sensor values have exceeded a predefined threshold.

**Case 3. Running of Python script to log on to edge gateway and repair malfunction:** A connected worker uses an iPhone mobile device to connect to the edge gateway over the SSH protocol and run Python scripts to repair a specific physical device malfunction, such as disabling a pump relay (i.e., LED), on site with no need for any manual action.

**Case 4. SCADA alarm and notification pop-up:** When a pump valve is opened or closed (i.e, Switch on/off LED) based on a decision received from a PLC or microcontroller, a message appears in the Web-SCADA interface.

*Case 5. Checking and monitoring of physical control systems on site:* The on-site supervisor uses an Android tablet to access the SCADA interface via the webSocket protocol and can also view real-time data graphs, current values, and the actuator's status using a mobile SCADA.

*Case 6. System monitoring:* The edge gateway provides local

SCADA/API systems over the webSocket protocol that enables an operator at the edge level to monitor the connected industrial physical systems and control them locally.

**Case 7. Maintenance of IIoT edge gateway**: The IT team can remotely (i.e., over the Internet) access the edge gateway using the SSH protocol to perform the processes required for technical maintenance.

**Case 8. Remote tuning of PLC parameters**: Operators at the enterprise level can access a Web-SCADA over the Internet (over HTTP protocol) using his/her authentication credentials and change the predefined values of the set-points of a PLC device

**Case 9. Ad hoc analytics for optimization**: The data analytics team can access a cloud application to determine how frequently a specific task related to the relevant sensor should be performed through searchable and filtered tables.

**Case 10. Fault modeling**: Through a cloud application, the data analytics team can create graphs for the sensor data collected during a specific period. Therefore, they can discover faults and any bias in the sensor values that help them identify any kind of malfunction that has happened or will occur.

**Case 11. Addition of new devices at edge**: The technical maintenance team can remotely add a new device to the IIoT edge gateway and configure it using the SSH protocol.

**Case 12. Malfunction in actuators**: If a specific actuator does not act according to the control command received, the operator will be able to notice this through a Web-SCADA system to quickly fix the problem remotely.

**Case 13. Quick query to edge data historian**: An operator sends a query to the data historian (i.e, a MySQL database in the edge gateway), asking about the last status of a specific actuator.

**Case 14. Remote starting or stopping of sensor**: An operator can start or stop a sensor from providing any measurement for local and cloud data historians. The simulated sensors and their engines are shown at the Web-SCADA/API interface where the operator can control them.

**Case 15. Backup and recovery**: A IT maintenance worker can access the edge devices, collect the desired logs and perform remote backup, and potentially provide a manual recovery process for connected industrial devices and edge gateways.

*D. Brown-IIoTbed Setup and Performance*

We implemented Brown-IIoTbed in the IoT lab of UNSW Canberra, which provides an isolated security testing environment. As shown in Fig. 3, the testbed setup consists of the aforementioned hardware, open-source software, and communications protocols (see Fig. 2). For example, Fig. 3 shows the installed Raspberry Pi and Arduino devices with their connected electronics components (i.e, sensors, and LEDs) which represent the physical asset and edge devices. Furthermore, it shows the management screen of ESXi 6.7 hypervisor which illustrates the WAN network's devices (e.g., Cloud application, and web-SCADA over HTTP), IT services script on the PC screen, laptop with running CoAP client scripts, and API on Android tablet.

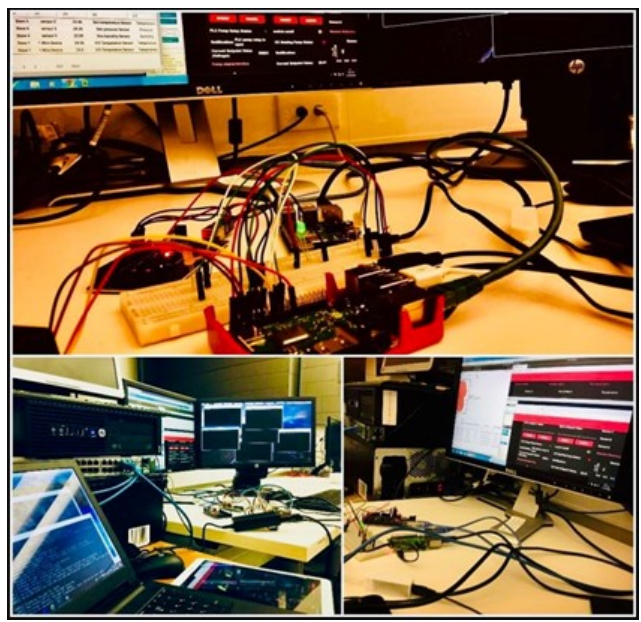

Fig. 3. Brown-IIoTbed setup

Brown-IIoTbed is successfully tested with connected wired and wireless devices. To evaluate its performance, we consider the edge gateway network and system activities as it is the central point of the Brown-IIoTbed at which the physical and cyber systems are connected. Fig.4 shows the edge gateway CPU load that is almost around 50% and it rarely peaks over 75% while Fig.5 shows that the system is making suitable use of memory where the total memory utilization is limited to less than 45%. Metrics related to the system's input/output activities are shown in Table I. As it can be observed, the workload of transfer, read, and write requests per second (their average values are 15.11, 6.13, and 8.98 respectively) is much low which in return has an efficient impact on the process response time. In overall, the edge gateway system (Raspberry pi 3 B+ with quad-core processor CPU) seems to work well. Fig.6 shows the average network throughput in Gigabit per second (Gbps) over a time interval (seconds). It can be seen that the average throughput drastically increases and its values are high. For example, the average throughput value at time interval 3600 seconds (i.e., the successfully transferred data rate in period 3000 to 3600 seconds) is approximately $5 \times 10^{+8}$ Gbps. Other network data analytic metrics are also shown in Table I such as average packet size, bytes/second, and bits/second have values 2153 bytes, 617BK/s, and 4941KB/s respectively.

To ensure the full connectivity and acceptable Quality of Service (QoS) for IIoT network, it is essential that the jitter, describing the variance in time delay between packets over time, does not exceed the certain tolerable value. This value varies in different applications; however, according to Cisco [38] the jitter value should be below 30ms and this value may roughly fit IIoT data streams. In our implementation, it can be clearly observed from Fig.7 that most of jitter values over time are expected and less than 30ms. In the light of these jitter values with existing wireless communication and the capabilities of edge gateway device (i.e., Raspberry pi 3 B+), complete" packets and it takes around 8.6ms in our implementation. WebSocket also starts out its connection by HTTP handshake, and then data is exchanged as frames over WebSocket protocol in persistent connection. Hence, the WebSocket response time is calculated based on HTTP, and it takes around 10.18ms. Arguably, this convenient level of service response times means low transport latency, appropriate network bandwidth utilization, and the reliable QoS.

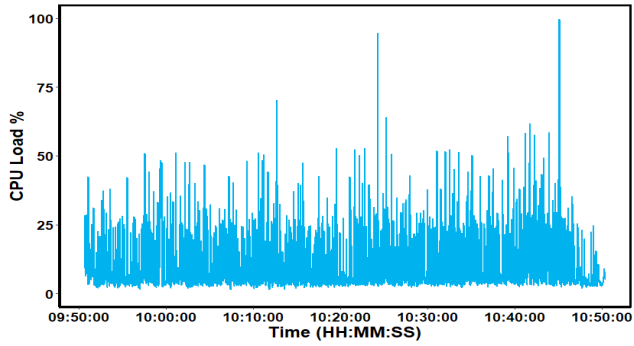

Fig. 4. Edge gateway system CPU Load

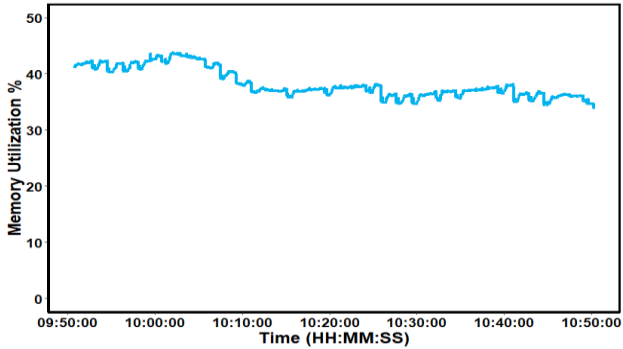

Fig. 5. Edge gateway system Memory Utilization

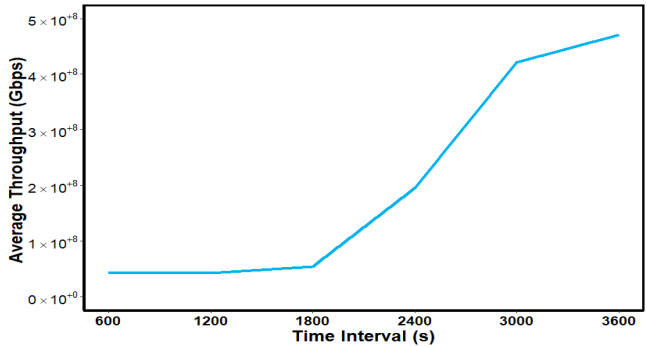

Fig. 6. Network Throughput over time interval

In summary, we can conclude that Brown-IIoTbed functions well and in a predicted manner. This is because Brown-IIoTbed is designed in a way that supports edge computing (using edge gateway), which provides faster process and storage and sends only the relevant data to the cloud. Further, it highly depends on emerging application protocols such as MQTT, CoAP, WebSocket that are designed to reduce network overhead, address latency and bandwidth problems, and provide QoS options.

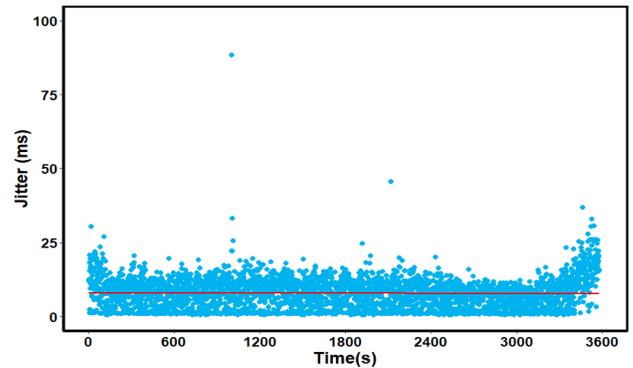

Fig.7. Network traffic Jitter over one hour

TABLE I
BROWN-IIOTBED PERFORMANCE METRICS OVER ONE HOUR

| | |
|---|---|
| Avg. Packet size (Byte) | 2153 |
| Avg. bytes/s (KiloByte) | 617 |
| Avg. bits/s (KiloByte) | 4941 |
| Avg. number of transfer request /second issued to physical device | 15.11 |
| Avg. number of read request /second issued to physical device | 6.13 |
| Avg. number of write request /second issued to physical device | 8.98 |
| Avg .MODBUS. Response time (ms) | 10.94 |
| Avg.CoAP. Response time (ms) | 7.38 |
| Avg.HTTP. Response time (ms) | 346.95 |
| Avg. DNS. Response time (ms) | 0.201 |
| Avg .I2C Response time (ms) | 1.34 |
| Avg. MQTT publish (QoS 2) message handshake time (ms) | 8.6 |
| Avg. SMTP response time (ms) | 12.3 |
| Avg.WebSocket.HTTP response time (ms) | 10.18 |

we can state that Brown-IIoTbed network performs well. A response time (i.e., sum of transport latency and processing time) is another important metric that has a bearing on QoS and needs to be retained under control. Table I shows the average of response time for the key Brown-IIoTbed application protocols. For example, the response time of MODBUS varies over the time but the average value is 10.94ms. This value fits the MODBUS TCP requirement where the recommendation for maintaining its value below 20ms [39]. While HTTP takes longer time around 346.95ms as it, in its way to the edge gateway, goes through router and firewall. However, it is still below the recommended limit value (1 second). Other services also provide acceptable response time including local DNS, CoAP, SMTP, and serial fieldbus I2C where take 0.201ms, 7.38ms, 12.3ms and 1.34ms respectively. While for MQTT protocol, as publish-subscribe protocol, we consider publishing (QoS=2) handshake time between client and broker (i.e., "publish message", "publish received", "publish release", and "publish complete" packets). It is the interval-time between "publish message" and "publish

*E. Examples of implementation of Brown-IIoTbed*

In this subsection, for the sake of brevity, we will provide examples of the testbed implementation of the most important communications, operations, and functionalities of Brown-IIoTbed.

The telemetry data collected from various physical devices is published via an MQTT protocol to the broker. Then, a

JavaScript function is used to parse the data, change its format, and send it as a JSON message to the cloud broker; for example, as shown in Fig. 8, a published packet with 225 bytes of an MQTT message and name [station/I2C slave] is sent from the edge gateway to the cloud broker. It consists of a JSON data representation that describes critical information, such as the device's ID/type and measurements. While the edge gateway polls the data from the MPL311A2 sensor using an I2C fieldbus protocol, parses and acts on received data format and then sends it in response to a connected client's request. This communication is conducted over the CoAP protocol between the CoAP client's mobile app (i.e., the connected worker) and the txThing CoAP server/edge gateway. The connected worker sends a 'GET' request to the resource path to read the sensor values and can also send a 'PUT' request to change the actuator's states (switch on/off). To perform this communication between a CoAP client and server, the original txThing code is updated and a new specific code for polling and acting on I2C sensor measurements is developed, as shown in Algorithm I.

Fig. 9 is a screenshot of the cloud application table for ad hoc analytics. The data collected at the cloud data historian can be accessed using this application and simultaneously filtered based on various factors. For example, the time-series data can be filtered based on the time interval between 2019-07-12 08:35:45.680 and 2019-08-09 10:41:30.000. Fig.10 shows the Node-Red flows created to connect a PLC with the edge gateway and other components, with the data register as the object on the right-hand side. The three registered addresses are received as voltage values on which the edge gateway can act and change the format to Celsius values (using a JavaScript code). This new data format can be passed to other system components, such as by being published to the cloud server using the MQTT client node.

Fig. 9. Data table in cloud application

Fig. 8. MQTT packets captured in Wireshark

Algorithm I. Generating COAP message
**Input:** *Request*
**Output:** *Response*
1. Data= Read *I2CDataBlock* (0x60, 0x00,6)
2. Var *temp_value* = Read *Temperature _ Register* ( )
3. *temp_value* = ((Data [4]*256) + (Data[5] & 0xF0) ) /16
4. CelsiusTemp = temp_value / 16.0
5. Var pressure_value= Read *Pressure_Register*( )
6. Pressure_value = (( Data[1]*65536) + (Data [2]*256) + Data [3] & 0xF0)) / 16
7. Pressure = (Pressure_value / 4.0 ) / 1000.0
8. Payload = String ({" Device Name"; "MPL3115A2", "data": {"Ctemp":{"Celsius:" CelsiusTemp}, "Pressure": {"Pascalpre": Pressure}}}
9. Response = CoAP.Message (code = COAP.Content, payload = Payload)
10. Return *Response*

## V. SECURITY TESTING AND ANALYSIS USING THE TESTBED

Each component of an IIoT system has a broad attack surface and may be a target for various cyberattacks. To prove the feasibility of the proposed testbed for security testing, the following attack scenarios are studied via experiments.

### A. Security threats based on STRIDE model

The Microsoft threat model 'STRIDE' (Spoofing (S), Tampering (T), Repudiation (R), Information disclosure (I), Denial of service (D) and Elevation of privilege (E)), the most common threat model, is adopted to describe various attack types. For the sake of brevity, only one attack scenario is tested for each attack type.

**Test Case 1: Spoofing attack (S)-Address Resolution Protocol (ARP) spoofing scenario**

In the ARP spoofing attack, the attacker sends faked ARP messages over a local network to link the attacker's MAC address with the IP address of the victim device [40]. In our implementation, the attacker can be between the edge gateway and router in the same network. The attacker begins to poison the edge gateway's ARP cache so that all the traffic from the edge gateway to the router passes via the attacker's machine. Fig. 11 shows an example of HTTP traffic before ARP spoofing where the Ethernet frame originates from the edge gateway with a MAC address of b8:27:eb:61:e5:14 to the router interface with a MAC address of 00:0c:29:6e:a7:ca. Fig. 12 shows the same HTTP traffic after ARP spoofing in which the frame has a different MAC address, that is, that of the attacker's machine of 00:0c:29:5b:a2:99. Through this process, the attacker can perform more malicious activities, such as preventing data from being sent to its destination or injecting false data or commands.

**Test Case 2: Tampering attack (T)-poisoning data analytics at cloud**

A tampering attack is based on the modification of data exchanged between the client and server [41]. In our implementation, this attack can occur between the cloud broker and edge gateway (as a publisher) where an attacker sends false data to the cloud to fake the sensor measurements. This injection is performed by a Man-in-the-Middle (MitM) attack that modifies the data-in-transit. Fig. 13 shows two superimposed curves in plots of pressure measurements over time. One is the true data collected from the local data historian at the edge gateway; the other is collected from the cloud data historian that is subject to tampering. The false data is fabricated such that it changes slowly to evade mechanisms for anomaly detection. This attack can continue for a long time as various data values are injected with the goal of affecting the decision-making process.

Fig. 10. MODBUS/TCP readings of Node-Red flows

Fig.11. Traffic from edge gateway to the router before ARP spoofing

Fig. 12. Traffic from edge gateway to the router before ARP spoofing

Fig. 13. Cloud data tampering

Fig. 14. Authorization Log

**Test Case 3: Repudiation attack (R)-sending fake notification and denying it by compromised edge gateway**

A repudiation attack happens when the user denies the fact that he/she has executed a specific action [42]. In our implementation, this can be performed by the edge gateway. The edge gateway often sends a notification to a connected worker which can be exploited by an attacker sending fake notifications (that are opposite to the real ones). When a worker realizes that an e-mail notification is not real, the edge gateway denies sending it. For example, we use the Node-Red library to send an e-mail to a connected worker, with each activity recorded in the Node-Red logs. The attacker can compromise the edge gateway and use a Python script to establish a connection to an SMTP server by harvesting credentials and e-mail accounts from the edge gateway. Afterward, the attacker can send a fake notification to the connected worker which will be denied by the edge gateway as this activity cannot be seen in the Node-Red logs. However, with adequate auditing and logs of the edge gateway (see Fig. 14), this activity can be proven to be conducted by the edge gateway using a Python script which indicates that it is due to either another legitimate user or a malicious one.

**Test Case 4: Information disclosure attack (I)-injecting edge gateway and sniffing fieldbus I2C protocol**

This type of attack aims to acquire specific information about the system [43]. For example, an attacker can monitor and capture the traffic between master and slave devices using sniffing scripts to understand the conversation between the endpoints and use the collected information for advanced attacks. Fig. 15 shows the traffic between the edge gateway (a master device) and the MPL3115A2 sensor (a slave device) captured by a sniffer (i.e., Python script), where '[' represents for start, Acknowledgment '+' or nor '-', address (7 or 10 bits), Read/Write bit (1=Read, 0= Write), 'xx' two hexadecimal characters for each data byte and ']' for a stop. For example, the second part of the message '[C0+01+ [C1+5C+84+70+ 17+F0+00-]' shows the read command issued by the master device to the slave's standard address (C1 (0x60 plus read bit)) for the 6 data bytes 5C, 84, 70, 17, F0 and 00. Using this data register's address and the data bytes, an attacker can extract the temperature values according to the formula (((Data [4]*256) + (Data [5] & 0xF0)) /16)16.0), that is, 23.9375 °C which could help the attacker to perform advanced threats such as a false data injection attack.

```
[C0+01+[C1+5C+84+70+17+F0+00-]
[C0+01+[C1+5C+84+E0+18+00+00-]
[C0+01+[C1+5C+84+90+18+00+00-]
```

Fig. 15. Sample of I2C traffic

**Test Case 5: DoS attack scenario (D)-attacking PLC device using Modbus/TCP protocol**

In an attack performed against PLC devices, the attacker with access to the network floods these devices with Modbus packets by sending huge numbers of read queries for various addresses of the targeted PLC [13]. The goal of this DoS attack can be to cause system disruption and expend processing resources. Fig. 16 shows the numbers of packets and transfer requests from a PLC to a connected physical I/O slave device during the DoS attack. It is clear that there is an abrupt increase in both these measurements. However, in this implementation, the PLC device still operates during and after the attack and proves its availability and resiliency.

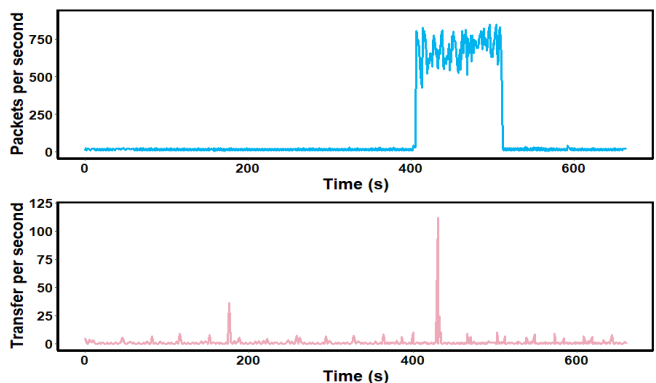

Fig. 16. Packets and transfer request/second during Modbus DoS attack

**Test Case 6: Elevation of privilege attack (E)-unprivileged subscriber to cloud MQTT broker**

An elevation of privilege attack occurs when an attacker obtains the rights of authorized users and gains a higher level of privilege to the system [44]. In our experiments, a malicious subscriber can connect to the cloud broker to obtain all the information exchanged between it and the publishers. This attack can be considered an elevation of privilege as this unauthorized subscription for several topics at the same time results in a disclosure of sensitive information exchanged between endpoints and the state of the broker. An MQTT subscriber for wildcard topics # and script responses, as shown in Fig. 17, displays the target broker in its raw format where the broker's 'Mosquito version 1.4.15', its states and the most recent published messages, such as the data received from a PLC Modbus, and simulated devices, are highlighted.

```
| $SYS/broker/subscriptions/count: 20
| $SYS/broker/load/bytes/sent/1min: 4921.41
|     station/PLC: {"Device ID":"Slave 2","Device Type":"PLC
MODBUS","Measurement":"16.407","Function":"PLC
Temperature Sensor","Content Type":"Temperature"}
….
| $SYS/broker/version: mosquitto version 1.4.15
| $SYS/broker/bytes/sent: 31348
| $SYS/broker/load/connections/5min: 1.32
| $SYS/broker/timestamp: Tue, 18 Jun 2019 11:42:22 -0300
|     station/sensor1: {"Device ID":"Slave 4","Device
Type":"sensor-1","Measurement":17.65,"Function":"Sim-
humditiy Sensor","Content Type":"Humditiy"}
| $SYS/broker/load/messages/received/15min: 27.11
| $SYS/broker/heap/current: 36688
```

Fig. 17. MQTT broker's subscribing data

### B. Machine Learning approaches for Intrusion detection

Utilizing Machine learning for analyzing and revealing malicious behavior is getting predominant as it can extract the hidden pattern of malicious and normal behavior [45]. Hence, we implemented machine learning approaches to identify the aforementioned attacks that have distinguishable features in network traffic (i.e., ARP spoofing, poisoning, Modbus DoS, and unprivileged MQTT subscriber). We utilized the most common algorithms including Random Forest (RF), Decision Tree (J48), Logistic Regression (LR), K-Nearest Neighbor (KNN), and Naive Bayes (NB). We collected and processed the raw network traffic of testbed during normal (without attack), and attacks using the Wireshark tool and then extracted the basic features of network traffic conversations. Examples of these features include the number of packets, number of bytes, bit rate, duration, and among others [46]. The final dataset has 42594, 145, 149, 1039, and 1534 conversations for normal, ARP spoofing, poisoning, Modbus DoS, and unprivileged MQTT subscriber respectively. We utilized the Accuracy (ACU), Precision (P), Recall (R)/detection rate, and F-Measure (F-M) based on 10-fold cross-validation to evaluate approaches' performance.

Table II shows machine learning approaches for intrusion detection where all of them achieved a reasonable performance. However, the RF achieved the best performance (i.e., 99.9%) in terms of accuracy, precision, recall, and F-Measure. While the NB achieved the worst performance compared with others because it assumes the independency among features, and it classifies the conversation based on the prior knowledge of the condition that may be related to the occurrence of conversations. Table III shows the performance of machine learning approaches in revealing attack types. As it can be observed, the RF achieved the highest detection rate for ARP spoofing (71.0%), and Modbus DoS (99.5%), while the NB obtained the best detection rate for the poisoning and unprivileged MQTT subscribers.

TABLE II
PERFORMANCE METRICS

| Approach | ACU (%) | P (%) | R (%) | F-M (%) |
|---|---|---|---|---|
| RF | **99.9** | **99.9** | **99.9** | **99.9** |
| J48 | 99.8 | 99.8 | 99.8 | 99.8 |
| NB | 42.4 | 95.4 | 42.3 | 55.1 |
| LR | 95.9 | 92.7 | 96.0 | 94.2 |
| KNN | 94.7 | 92.6 | 94.7 | 93.6 |

TABLE III
DETECTION RATE FOR ATTACKS (%)

| Approach | ARP spoofing | Modbus DoS | poisoning | Unprivileged MQTT Subscriber |
|---|---|---|---|---|
| RF | **71.0** | **99.5** | 96.6 | 99.7 |
| J48 | 65.5 | **99.5** | 96.0 | 99.7 |
| NB | 3.4 | 89.0 | **98.7** | **99.9** |
| LR | 3.4 | 99.0 | 44.3 | 0.0 |
| KNN | 64.1 | 98.7 | 4.7 | 1.0 |

*C. Security evasion / Malicious reverse shell backdoor against Router/Firewall for remote command injection*

Edge devices are usually protected from the external world by firewalls that specify the IP addresses that can legitimately access their internal networks. However, attackers are always able to pass such security mechanisms by identifying their weaknesses, such as misconfigurations, and unpatched vulnerabilities. In our experiments, a pfsense router/firewall with vulnerability (CVE-2016-10709) is used. An attacker starts its exploitation process by scanning and numerating target devices to identify open ports and collect more system information. For instance, the output of Nmap command (Nmap -f –v –O 192.168.10.1) is "PORT STATE SERVICE 443/tcp open https and OS details: FreeBSD 7.0-RELEASE-p1 - 10.0-CURRENT, FreeBSD 7.2-RELEASE", which indicates the possibility of accessing the WebGUI of the router/firewall device. An attacker can use the default credentials or perform a phishing attack to gain access to it and find any vulnerability that can be exploited. Another way that can be performed is a directory and file traversal (web application attack) e.g., using the Dirbuster tool against (https://192.168.10.1:443), to find files that may have information related to these credentials. Based on the collected credentials and information, the attacker can exploit the pfsense software's vulnerability by injecting malicious PHP codes and creating reverse shell backdoor for command injection with root privileges, as shown in Fig.18.

*C. Malicious payload hunting and intelligence*

As attacks have evolved and attackers keep working on new techniques and tactics to compromise systems, more techniques than prevention and reactive detection are required

```
msf exploit(unix/http/pfsense_graph_injection_exec) > exploit
[*] Started reverse TCP handler on 192.168.10.151:4444
[-] pfSense version not detected or wizard still enabled.
[*] Payload uploaded successfully, executing
[*] Command shell session 1 opened (192.168.10.151:4444 -> 192.168.10.1:64609) at 2019-10-01 22:38:42 -0500
```

Fig. 18. Example of Metasploit command to exploit pfsense vulnerability

to reveal their behaviors. Threat hunting [47], which chases an attacker (rather than being a target and working in the passive mode), is a reasonable solution. The hunting process for malicious reverse shell backdoor payload (see Section V- B) starts by verifying the main hypothesis that "an attacker may be operating on a reverse shell backdoor that uses to launch malicious commands with root privileges to the router/firewall". To verify this hypothesis, it is necessary to collect data from multiple sources and, given basic knowledge of the target device and network, the indicators may be found in the abnormal network's connections and system log of router/firewall.

Firstly, given the network connection data obtained from the Zeek tool (formally known as Bro) [48], all the client IP addresses connected to a pfsense router/firewall WebGUI using open HTTPS port ( i.e., 443) can be extracted. The Bro connection logs (i.e, conn.log) that fit an Excel file are used to search for connected IP addresses and their accumulated traffic statistics such as total connections' duration and number of data bytes sent. Shown in Fig. 19 are all the IP addresses that start connections to the router/firewall on port 443, total number of connections (i.e., count), total connections' duration (i.e., duration is the difference in time between the first and last packets seen), maximum and minimum durations, and total number of sending bytes that are extracted from the determined connections. As the IP address 192.168.10.151 has a suspicious behavior that is the longest connection time and most data bytes sent, the question is "Did the router/firewall connect back to this IP address as this might be an indicator for reverse shell?". To answer this, new queries that return the connections that may have originated from the router/firewall back to one of the aforementioned IP addresses are performed. As shown in Fig. 20, the router/firewall establishes a connection back to the IP address 192.168.10.151 five times on port 4444 with a total duration of 1344.026 seconds. By tracking the TCP data streams listener on that port (i.e., 4444) and checking type of exchanged packets between victim IP (i.e, 192.168.10.1) and suspicious IP (i.e., 192.168.10.151), we found that most of exchanged packet types are [PSH, ACK], and [ACK] TCP packets. This information indicates the possibility of a reverse shell backdoor that is opened back from the router/firewall to an attacker's machine on port 4444 many times over TCP protocol.

The next step in malicious reverse shell backdoor payload hunting process is identifying the malicious payload, and this step is accomplished by analyzing the system logs after collecting, extracting and parsing them into two columns each with a timestamp and event in a CSV file, and searching for long malicious commands launched by attackers. As shown in Fig. 21, a Python script is developed to parse the data and filter the longest commands (more than 250), with the search yielding a positive finding. Examining the logs, the longest malicious command with sensitive characters and numbers (i.e., 1004) is identified as a malicious PHP script injected into the router/firewall WebGUI whereby the attacker uses a known vulnerability (unpatched by the system administrator) in status_rrd graph_img.php to inject an obfuscated malicious code into a 'throughput-rrd.file-printf' command. To obtain more information about this payload, it is de-obfuscated and fitted as a text file to the VirusTotal [49] which has multiple anti-virus tools for detecting malicious files. As shown in Fig. 22, VirusTotal has a high false negative rate, with only one of its 56 anti-viruses, BKav, identified the text file as a web shell while most of the anti-virus tools failed to detect a malicious payload (Metasploit payload) as they do not have a signature related to this malicious payload. This indicates the importance of hunting rather than using only detection engines.

| X | Y | Z | AA | AB | AC |
|---|---|---|---|---|---|
| Source IP Address | Count | Total Duration | Max_Duration | Min_Duration | Total origniated bytes |
| 192.168.10.151 | 31 | 1614.25897 | 345.580034 | 0.002894 | 174325 |
| 192.168.10.155 | 34 | 915.820542 | 140.98219 | 0.007694 | 131690 |

Fig. 19. List of connections established with router/firewall over port 443

| Z | AA | AB | AC | AD |
|---|---|---|---|---|
| Source IP address | Destination IP address | Destination port | Count | Total Duration |
| 192.168.10.1 | 192.168.10.151 | 4444 | 5 | 1344.026 |

Fig. 20. List of connections established back from router /firewall

In summary, the network indicators learned from the first step, the HTTPS connections over open port 443 with the router/firewall are leveraged as the starting point for tracking suspicious behaviors. Finding the longest duration, the largest number of originated bytes and a reverse connection of the server to the specific port of a suspicious IP for a long time is useful for tracking suspicious TCP stream data. Also, filtering the longest commands in the logs helps to detect a malicious reverse shell backdoor payload for remote command injection. All the stacked information confirms the proposed hypothesis.

Fig. 21. Parsing and filtering system logs

Fig. 22. Results obtained from VirusTotal

## VI. COMPARISONS AND DISCUSSION

To compare our new testbed with existing ones, the following features are used [17, 21, 50]:

- **Usability:** A testbed should be easy to use, learn, configure, build, operate and reproduce. Also, understanding its scenarios and interpreting its output should be simple.
- **Fidelity:** The design of a testbed should follow an agreed international standard architecture, that is, an IIRA model for IIoT systems (see Section III). This design expects to cover the IIoT system's main components and functionalities. This feature can also focus on interoperability, and closed control loop as important characteristics of brownfield IIoT systems.
- **Heterogeneity:** A testbed should have different physical access media (i.e., wired and wireless), application and industrial fieldbus protocols (i.e., CoAP, MQTT, HTTP, WebSocket, Modbus/TCP, I2C, 1-wire), various devices (i.e., sensors and physical control devices), API/applications and web browser interfaces (i.e., cloud, mobile applications and among others).
- **Flexibility and Scalability**: A testbed should be able to be modified, changed, and expanded, including being adaptable, sustainable and customizable. For example, new specific-environment sensors over specific communications should be involved at the edge layer to connect with the edge gateway. Other devices, simulators, and applications should be also included.
- **Federation:** A testbed should offer various experimental capabilities on the same standardized platform so that experiments can be repeated.
- **Safety, Reliability and Resilience:** A testbed should support an industrial control system's characteristics of safety, reliability, and resilience. Examples of a safe scenario need to be designing a physical control system with specific parameters for sending alarms and notifications. The system should behave predictably and use protocols and a fieldbus that specify safety standards, such as counters, time-outs, unique sender and receiver identifications, and cross-checks. Reliability is a testbed's capability to perform the required functions under the stated conditions for a specific time interval: For example, the actuators are programmed to react to the physical environment, i.e., switch on/off for specific times whenever the relevant condition is met. Resilience is a testbed's capability to absorb any incident and continue working without significant effects: For example, there should be a recovery and backup procedure for the collected data. Also, an IIoT system should be supported by the implementation of specific techniques: For instance, in the case of a crash, the sensor, controller, and other devices should still work.

- **User Interfacing**: Simple tools should be available to fix a testbed's malfunction, change its configuration, and support its programming and logging functionalities: For example, using SSH connectivity for all system's devices.
- **End-to-End testbed:** A testbed should be holistic and end-to-end to provide three layers of the IIoT system: edge layer (i.e. physical devices and edge computing), platform layer (i.e. cloud storage, and analytics), and enterprise layer (i.e. service and application devices).
- **Primary purpose:** The key objective is to develop a testbed with certain levels of specialization, such as security testing and application for IIoT systems.

Based on these features [17, 21, 50], an analysis of existing IoT/IIoT testbeds is conducted, with the comparative results shown in Table IV. It is found that most of the existing testbeds [15, 17, 20, 21, 23, 24] focused on general IoT system implementations rather than industrial ones and, at most, satisfied 7 features for IIoT testbeds, with only one specific for IIoT 'INFINITE' [19]. The 'INFINITE' testbed has been built by the IIC and is still under development. It was not designed for security testing and there is little public information available about it. However, our testbed (Brown-IIoTbed) achieved 13 of the relevant features.

Implementing a holistic end-to-end IIoT system is considered a highly complex task due to the need to integrate Operational Technology (OT) systems (i.e., hardware and software interact with the physical process) and IT systems, and different functionalities and processes (from the edge and cloud to services and applications) with maintaining key characteristics of an IIoT system of safety, reliability, and resilience. Moreover, the need to support the system's heterogeneous nature where there is a wide variety of M2M, M2H, and H2M communications, various devices, access media, APIs, and states. Brown-IIoTbed deals with these challenges by developing a generic, affordable, and high-fidelity end-to-end IIoT testbed that provides the key functionalities in a simple and easily understandable manner. It utilizes cost-effective devices, free open-source software, and affordable computer devices to provide a testbed that is scalable, adaptable, and can be reproduced, modified, and changed to fit the research demand. Brown-IIoTbed covers existing gaps in the development of IoT/IIoT system testbeds by supporting various new IIoT application protocols, legacy industrial protocols, the interoperability among them, and providing edge computing, and various APIs for visualizations of data analytics, controlling, and monitoring physical assets.

Moreover, Brown-IIoTbed deals with one of the greatest challenges in the IIoT research, that is, security testing. IIoT security (in particular for brownfield) is considered an extremely complex process as IIoT systems integrate both OT and IT technologies, which have different security perspectives and priorities of data protection, physical process, and control. Addressing and analyzing the security issues related to such implementations need to be well understood for the purpose of identifying potential threats and targets, and estimating the possible consequences. This is achieved in our proposed testbed by using a STRIDE model to provide and analyze examples of potential threats against various IIoT system components, and machine learning approaches for intrusion detection. An advanced security testing, vulnerability exploitation, and malicious payload injection against the router/firewall are presented. Threat hunting and intelligence related to a malicious reverse shell payload as an example of proactive defense techniques are also performed in Brown-IIoTbed. Our testbed can be also extended with other security tests that fulfill researchers' demands.

However, Brown-IIoTbed has several limitations. For example, since IIoT implementations are still in their early stages and most of the existing implementations are special projects rather than standard and publicly-available ones, it is not possible to make a comparison between Brown-IIoTbed performance and real systems. We cannot show that the Brown-IIoTbed behavior is similar to the real system and this is a key limitation for our work. Nevertheless, we can state that it has been solved somewhat by adopting IIRA as a standard reference model to build a high fidelity testbed. But in the meantime, this remains a limitation. Additionally, Brown-IIoTbed, like the most advanced testbeds, has limited ability in simulating real environmental conditions for process measurement and control systems. Another limitation of Brown-IIoTbed is the limited edge gateway hardware capabilities (i.e., Raspberry pi). If an experiment needs to connect more PLC devices to the edge gateway, this may cause an increase in the load of the edge gateway and affect its performance. To expose this limitation, another edge gateway can be used to connect these PLC devices, and then the two-edge gateways can be connected together. This will act like a traditional Distributed Control System (DCS) [12] where many master devices connect to each other. A further solution can be achieved by connecting two-edge gateways separately and directly to the cloud.

## VII. Conclusion and Future work

This paper presents a new IIoT testbed that enables security researchers to easily reproduce it and test their security hypotheses while also facilitating analyses of potential attacks, and attackers' techniques and tactics. The proposed testbed (i.e, Brown- IIoTbed) provides generic, holistic, high- fidelity and simplified end-to-end instance of realistic IIoT systems at a lower cost without requiring any additional management, maintenance, high-level skills or domain expertise, and without causing real systems any physical risk or damage. To the best of our knowledge, this is the first end-to-end IIoT testbed that is developed for security testing with the main focus on Brownfield implementation. We demonstrate the feasibility of Brown-IIoTbed in conducting security testing by providing various attacks based on a STRIDE threat model,

TABLE IV
COMPARISON OF EXISTING TESTBEDS

| Testbed | U | F | | H | | | | | FL and SC | FE | S, REL and RES | UI | E2E | Primary Purpose |
|---|---|---|---|---|---|---|---|---|---|---|---|---|---|---|
| | | IIRA | I | CCL | AIP | D | PAM | API/A | | | | | | |
| Patel et al.[15] | ● | ○ | ○ | ○ | ○ | ● | ○ | ◉ | ◉ | ● | ○ | ○ | ○ | Testing IoT application |
| Choosri et al. [16] | ● | ○ | ○ | ○ | ○ | ● | ○ | ◉ | ◉ | ● | ○ | ○ | ○ | Smart Traffic light |
| Deshpande et al. [17] | ● | ○ | ○ | ● | ○ | ● | ○ | ○ | ● | ● | ○ | ○ | ○ | Industrial automation |
| Merchant and Ahire [18] | ● | ○ | ○ | ● | ○ | ● | ○ | ◉ | ● | ● | ○ | ○ | ○ | Industrial automation |
| INFINITE [19] | ◉ | ● | - | ● | - | ● | ● | - | ● | ● | ● | - | - | Testing new IIoT devices/ technologies |
| FIESTA-IoT [20] | ◉ | ○ | ○ | ◉ | - | ● | ● | ○ | ● | ● | ○ | - | ○ | Testing IoT applications |
| JOSE [21] | ◉ | ○ | ○ | ○ | ○ | ● | ● | ○ | ● | ● | ○ | ● | ○ | IoT service evaluation |
| Siboni et al. [22] | ● | ○ | ○ | ◉ | ○ | ● | ● | ○ | ● | ● | ◉ | ● | ○ | IoT security testing |
| Berhanu, Abie, and Hamdi [23] | ● | ○ | ○ | ○ | ○ | ● | ● | ○ | ◉ | ● | ○ | ● | ○ | Testing healthcare-IoT security |
| Hossain et al. [24] | ◉ | ○ | ○ | ○ | ○ | ● | ● | ○ | ● | ● | ○ | ● | ○ | Generic IoT experiments |
| Brown-IIoTbed | ● | ● | ● | ● | ● | ● | ● | ● | ● | ● | ● | ● | ● | IIoT security testing |

**U**: **U**sability    **F**: **F**idelity    **H**: **H**eterogeneity    **IIRA**: **I**ndustrial **I**nternet **R**eference **A**rchitecture
**CCL**: **C**losed **C**ontrol **L**oop    **AIP**: **A**pplication and **I**ndustrial **p**rotocol    **UI**: **U**ser **I**nterfacing    **PAM**: **P**hysical **A**ccess **M**edia
**API/A**: **A**pplication **P**rogramming **I**nterface /**A**pplication    **FL and SC**: **Fl**exibility **and Sc**alability    **FE**: **Fe**deration    **D**: **D**evices
**S, REL, and RES**: **S**afety, **Rel**iability, **and Res**ilience    **E2E**: **E**nd to **E**nd testbed    **I**: **I**nteroperability

●: Features considered    ○: Features not considered    ◉: Features not explicitly and completely considered    - : not available information and evasion security/reverse shell backdoor against router/firewall. We also utilize Brown-IIoTbed to provide machine learning approaches for intrusion detection, and perform malicious payload hunting and intelligence as a proactive defense technique (early detection). The results show that Brown-IIoTbed is well-structured and performed satisfactorily for security testing and operated in a way that will enable researchers to study various security issues. Also, in-depth analyses and comparisons with existing IoT/IIoT testbeds are conducted. The overall results prove that Brown-IIoTbed satisfied the 13 features required for an IIoT testbed (described in section IV-Table I) and cover the research gaps regarding existing testbeds.

In future work, we plan to improve this testbed's capacity and performance by deploying more and different sensors, machines, industrial protocols, and applications, implementing complex ladder logic programs for PLCs. We will also expand the capabilities of cloud applications by performing big data analytics using deep and machine learning techniques. Also, we intend to generate an intrusion dataset for security research purposes.